\documentclass[twocolumn,showpacs,preprintnumbers,amsmath,amssymb]{revtex4}
\usepackage{graphicx}

\begin{document}
\def\be{\begin{equation}} 
\def\ee{\end{equation}}
\def\bearr{\begin{eqnarray}}
\def\eearr{\end{eqnarray}}
\def\tc{$T_c~$}
\def\tcl{$T_c^{1*}~$}
\def\c2{ CuO$_2~$}
\def\ruo{ RuO$_2~$}
\def\lsco{LSCO~}
\def\bi{bI-2201~}
\def\tl{Tl-2201~}
\def\hg{Hg-1201~} 
\def\sro{$\rm{Sr_2 Ru O_4}$~}
\def\rc{$RuSr_2Gd Cu_2 O_8$~}
\def\mgb{$MgB_2$~}
\def\pz{$p_z$~}
\def\ppi{$p\pi$~}
\def\sqo{$S(q,\omega)$~}
\def\tperp{$t_{\perp}$~}
\def\he4{${\rm {}^4He}$~}
\def\ags{${\rm Ag_5 Pb_2O_6}$~}
\def\nxcob{$\rm{Na_x CoO_2.yH_2O}$~}
\def\lsco{$\rm{La_{2-x}Sr_xCuO_4}$~}
\def\lco{$\rm{La_2CuO_4}$~}
\def\lbco{$\rm{La_{2-x}Ba_x CuO_4}$~}
\def\half{$\frac{1}{2}$~}
\def\thalf{$\frac{3}{2}$~}
\def\tst{${\rm T^*$~}}
\def\tch{${\rm T_{ch}$~}}
\def\jeff{${\rm J_{eff}$~}}
\def\nbc{${\rm LuNi_2B_2C}$~}
\def\cabc{${\rm CaB_2C_2}$~}
\def\nboo{${\rm NbO_2}$~}
\def\voo{${\rm VO_2}$~}
\def\nip{$\rm LaONiP$~}
\def\nisb{$\rm LaONiSb$~}
\def\nibi{$\rm LaONiBi$~}
\def\fep{$\rm LaOFeP$~}
\def\cop{$\rm LaOCoP$~}
\def\mnp{$\rm LaOMnP$~}
\def\fesb{$\rm LaOFeSb$~}
\def\febi{$\rm LaOFeBi$~}
\def\efeas{$\rm LaO_{1-x}F_xFeAs$~}
\def\hfeas{$\rm La_{1-x}Sr_xOFeAs$~}
\def\hSfeas{$\rm Sm_{1-x}Sr_xOFeAs$~}
\def\hCefeas{$\rm Ce_{1-x}Sr_xOFeAs$~}
\def\feas{$\rm LaOFeAs$~}
\def\Ndfeas{$\rm NdOFeAs$~}
\def\Smfeas{$\rm SmOFeAs$~}
\def\Prfeas{$\rm PrOFeAs$~}
\def\refeas{$\rm REOFeAs$~}
\def\refesb{$\rm REOFeSb$~}
\def\refebi{$\rm REOFeBi$~}
\def\ttog{$\rm t_{2g}$~}
\def\fese{$\rm FeSe$~}
\def\fete{$\rm FeTe$~}
\def\eg{$\rm e_{g}$~}
\def\dxy{$\rm d_{xy}$~}
\def\dzx{$\rm d_{zx}$~}
\def\dzy{$\rm d_{zy}$~}
\def\dxsq{$\rm d_{x^{2}-y^{2}}$~}
\def\dzsq{$\rm d_{z^{2}}$~}

\title{Exact quantum spin liquids with Fermi surfaces in spin-\half models}

\author{ G. Baskaran, G. Santhosh and R Shankar}
\affiliation
{The Institute of Mathematical Sciences,
C.I.T. Campus, Chennai 600 113, India }

\begin{abstract}
An emergent Fermi surface in a Mott insulator, an exotic quantum spin liquid state, was suggested by Anderson in 1987. After a quick support for its existence in spin-\half Heisenberg model in a square lattice in a RVB mean field theory, pseudo Fermi surface was found only recently in an exactly solvable spin-$\frac{3}{2}$ model by Yao, Zhang and Kivelson. We show that a minimal spin-\half Kitaev model on a decorated square lattice exhibits a Fermi surface. Volume and shape
of the Fermi surface change with exchange couplings or on addition of a 3 spin interaction terms.
\end{abstract}

\maketitle
Fermi surface emerges as a remarkable consequence of Pauli exclusion principle. It is a non-trivial organization of \textit{itinerant fermions} in their ground states. So it has a natural place in describing a metal, where there are itinerant electrons. However, it was conjectured by Anderson\cite{pwaScience}, that even in the absence of itinerant electrons a \textit{pseudo Fermi surface} could arise from \textit{itinerant spins} in a Mott insulator. Such a possibility was soon confirmed in a resonating valence bond (RVB) mean field theory\cite{bza} for a spin-\half Heisenberg antiferromagnet on a square lattice. In an exciting development, Kitaev\cite{KitaevAnnPhys} constructed a spin-\half model on a honeycomb lattice, where gapless noninteracting fermionic excitations emerge in a most natural fashion as an exact solution. There is no Fermi surface however, only a Fermi point. It is an important question if one can change the model suitably and get a Fermi surface enclosing a finite volume of the Brillouin zone; and whether one can cause shape and volume change of the Fermi surface by tuning coupling parameters.

On the experimental front, after years of efforts\cite{PatrickLee}, certain spin-\half Mott insulating organic ET salts\cite{kanoda}, and Na$_{4}$Ir$_{3}$O$_{8}$\cite{takagi}, 
a 3 dimensional hyper kagome lattice antiferromagnet, exhibiting  quantum spin liquid containing a pseudo Fermi surface have been synthesized. The spin Hamiltonian for the organic insulators, unlike the model studied in the present paper, has a SU(2) global spin rotational symmetry. The present model, which does not have a global SU(2) symmetry, has however features that should teach us about these real systems and systems describable by Kugel and Khomskii quantum compass model\cite{khomskii}.

\begin{figure}[t]
\begin{center}
\includegraphics[width=3in]{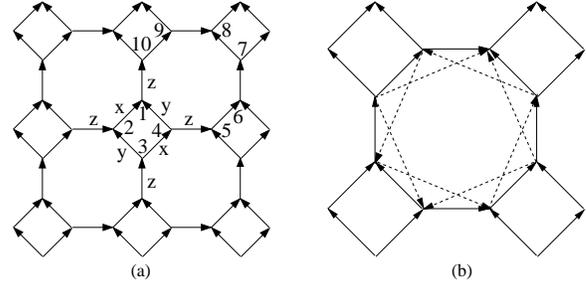}
\caption{(a) decorated square lattice (Fisher lattice) with zero-flux in each plaquette. The gauge variable $u_{ij}$ (see text) is $+1$ if the arrow in the bond $<ij>$ points from $i$ to $j$. (b) dotted lines represent 3-spin interactions and the arrows on them represent the values of the variable $Z_{ij}$ (see text).}
\label{FigLattice}
\end{center}
\end{figure}

We show exactly that spin-\half Kitaev model defined on a suitably decorated square lattice
(Fisher lattice) possesses the desired psuedo Fermi surface in certain flux sector. We consider the translationally invariant model containing three types of couplings J$_{x}$, J$_{y}$ and J$_{z}$.
We get a well defined fermi surface at the isotropic point: J$_{x}$ = J$_{y}$ = J$_{z} = J$.
As we move away from this point the shape and volume of the Fermi surface change and eventually disappear. A 3 spin interaction term also  induces change in shape and volume of the Fermi surface, leading to an eventual disappearance of the Fermi surface. In a recent work Yao, Zhang and Kivelson (YZK)~\cite{YaoZhangKiv} have suggested a spin-$\frac{3}{2}$ model on a square lattice where they find pseudo Fermi surface for a range of parameters. Nussinov and Ortiz~\cite{NussinovOrtiz} also discussed a class of models, whose spin representations include a spin-$\frac{1}{2}$ system on a square lattice bilayer, with Fermi surface. These models may be viewed as non-trivial generalisation of the Kitaev model. The merit of our work is that even the simpler spin-\half Kitaev model, defined on a decorated square lattice(Fig.~\ref{FigLattice}) leads to a pseudo Fermi surface. The advantage of our construction is that it allows for possibility of Fermi surface even in certain three dimensional lattices~\cite{BaskaranSanthoshShankar}. Kitaev model on a variety of interesting lattices was studied by~\cite{YangZhouSun}.

Ordered quantum antiferromagnetic states and their elementary excitations are well described by standard spin wave theories. However, these semiclassical theories are not capable of describing a quantum spin liquid state and their elementary excitations. This is because there is no manifest long range order describable by standard order parameters. It is here, RVB mean field theory broke new grounds in 1987 and introduced methods that were novel and unconventional. It is also interesting to note that RVB mean field approach, though yields approximate results for isotropic Heisenberg antiferromagnet in 2D, reproduces exact results for the Kitaev model\cite{ShankarBaskaran}, an anisotropic spin model.

In view of this we briefly review RVB mean field theory of pseudo Fermi surface of spin-\half Heisenberg antiferromagnet on a square lattice.  This theory differs from Anderson-Fazekas'~\cite{AndersonFazekas} variational approch and was more ambitious: it developed a new formalism to study quantum spin liquid states and their excitations. It involved a key step of enlarging the Hilbert space by writing the spin operators in terms of electron operators constituting the spin half magnetic moment and introducing bond singlet operators. In this representation the Heisenberg Hamiltonian takes the form,
\be
 H_s = J\sum_{\langle ij\rangle} ({\bf S}_i \cdot {\bf S}_j - \frac{1}{4}) \equiv  -J
 \sum_{\langle ij \rangle} b^{\dagger}_{ij}b^{}_{ij}
\ee
using the relation  ${\bf S}_i \equiv \sum_{\alpha, \beta} c^{\dagger}_{i\alpha}
{\vec{\tau}}_{\alpha\beta} c^{}_{i\beta}$, where ${\tau^{\alpha}}$ are the Pauli spin matrices and $c$'s are the electron opertators that constitute the physical magnetic moment. Further $b^{\dagger}_{ij} \equiv \frac{1}{\sqrt2} (c^{\dagger}_{i\uparrow}c^{\dagger}_{j\downarrow}-
c^{\dagger}_{i\downarrow}c^{\dagger}_{j\uparrow})$ is the bond singlet operator. In the electron representation we have an enlarged Hilbert space of dimensions 4$^{N}$. The local constraint, $n_{i\uparrow} + n_{i\downarrow} = 1$, reduces 4$^{N}$ to 2$^{N}$, the dimension of the physical Hilbert space.
The Hamiltonian enjoys a local U(1) symmetry because the local electron number operator $n_{i} \equiv n_{i\uparrow} + n_{i\downarrow}$ commutes with the Hamiltonian: [H,n$ _{i}$] = 0. Thus sectors containing same total number of particles $N$, but with $N-2M$ singly occupied sites, $M$ doubly occupied and $M$  unoccupied sites ($M$ = 0, 1, 2, ... $ \frac{N}{2}) $are superselected. Unlike the Kitaev Model, 
enlarging the Hilbert space does not allow for exact solvability of the isotropic Heisenberg antiferromagnet. However it suggests important approximation methods and solutions.

A Bogoliubov-Hartree-Fock factorization leads to a mean field Hamiltonian:
\be
 H_{\rm mF} =  -J \sum_{\langle ij \rangle}( \langle b^{\dagger}_{ij}\rangle b^{}_{ij} +
\langle b^{}_{ij}\rangle b^{\dagger}_{ij})
\ee
and simple mean field solutions $\langle b^{}_{ij}\rangle = \Delta_0$. After diagonalising the mean field Hamiltonian we get the quasiparticle Hamiltonian:
\be
H_{\rm mF} \sim J \sum_{k\alpha} 
|(\cos k_x + \cos k_y)|~\alpha^{\dagger}_{k\sigma} \alpha^{}_{k\sigma}
\ee
The quasi particle energy vanishes on a square connecting points $(\pm\pi,0), (0,\pm \pi)$ in the BZ. This defines the pseudo Fermi surface for spinon excitations. Unlike Kitaev model, the RVB mean field analysis of the Heisenberg Hamiltonian did not lead to 2$^{N}$ gauge copies with identical energy spectrum. So one needs to Gutzwiller project the mean field ground state to the physical subspace containing only single occupancies and then calculate the ground state and low energy properties.

Now we study Kitaev model on the decorated square lattice (Fig.~\ref{FigLattice}).  Kitaev model Hamiltonian is:
\be
\label{ham1}
H=-J_x\sum_{\langle ij\rangle_x}\sigma^x_i\sigma^x_j
-J_y\sum_{\langle ij\rangle_y}\sigma^y_i\sigma^y_j
-J_z\sum_{\langle ij\rangle_z}\sigma^z_i\sigma^z_j
\ee
where ${\langle ij\rangle_x}, {\langle ij\rangle_y}, {\langle ij\rangle_z}$ are the x, y and z-type nearest neighbor bonds shown in Fig.~\ref{FigLattice}. There are two types of elementary plaquettes 
(square and octagon) in the decorated square lattice (Fig.~\ref{FigLattice}). There are two types of local conserved operators, ${\rm S}_{p} = \sigma^z_1\sigma^z_2\sigma^z_3\sigma^z_4$ for the squares and ${\rm O}_{p} = \sigma^x_1\sigma^x_4\sigma^y_5\sigma^y_6\sigma^x_7\sigma^x_8\sigma^y_9\sigma^y_{10}$ for the octagons. The plaquette operators ${\rm S}_{p}$ and ${\rm O}_{p}$ commute among themselves and with Kitaev Hamiltonian.

To solve for the spectrum of the Hamiltonian, we follow Kitaev and introduce 4 types of Majorana fermions on each sites,
$c^\alpha,~\alpha=0,x,y,z$, defined by the anticommutation relations
$\{c^\alpha,c^\beta\}=2\delta_{\alpha\beta}$.
Four Majorana (real) fermions make two complex fermions, making the Hilbert 
space 4 dimensional. This is similar to enlargement of Hilbert space in RVB mean field theory. In fact, $c^{\alpha}$'s could be combined to form two complex fermions; $c_{\uparrow}=(c^0+ic^x)/2$,  $c^{\dagger}_{\uparrow}=(c^0-ic^x)/2$, $c_{\downarrow}=(c^y+ic^z)/2$, $c^{\dagger}_{\downarrow}=(c^y-ic^z)/2$.

Total number of spins in our system is $4N$, where $N$ is the number of unit cells. The dimension of Hilbert space of 4N spins is $2^{4N}$. The enlarged Hilbert space has a dimension 4$^{4N}$ 
$=(\sqrt{2}\times\sqrt{2}\times\sqrt{2}\times\sqrt{2})^{4N}$. Hereafter we denote $c_i^0$ by $c_i$. State vectors of the physical Hilbert space satisfy
the condition,
\begin{eqnarray}
\label{cons1}
D_i\vert\Psi\rangle_{\rm phys}=\vert\Psi\rangle_{\rm phys}, \quad
D_i \equiv c_i~c^x_i c_i^y c_i^z
\end{eqnarray} 
The spin operators can then be represented by,
$\sigma^a_i=ic_ic_i^a$ where $a\in \{x,y,z\}$.
When projected into the physical Hilbert space, the operators defined
above satisfy the algebra of spin $1/2$ operators,
$[\sigma^a_i,\sigma^b_j]= 2 i\epsilon_{abc}\sigma^c \delta_{ij}$.
The Hamiltonian written in terms of the Majorana fermions is,
\begin{eqnarray}
\label{KitaevHamiltonian}
H=-\sum_{a=x,y,z}J_a\sum_{\langle ij\rangle_a}~ic_i{\hat u}_{\langle ij\rangle_a}c_j, 
\end{eqnarray}
with ${\hat u}_{\langle ij\rangle_a}\equiv ic_i^ac_j^a$.
Following Kitaev we find that $ [H,{\hat u}_{\langle ij\rangle_a}]=0 $ and $u_{\langle ij \rangle a}$ become constants of motion with eigen-values ${u}_{\langle ij\rangle_a} = \pm 1 $. The variables ${u}_{\langle ij\rangle_a}$ are identified with static (Ising) $Z_2$ gauge fields on the bonds. As ${{\hat u}}_{\langle ij\rangle_a} = - {{\hat u}}_{\langle ji\rangle_a}$ we follow a definite ordering of the indices $ij$ in specifying the value of ${u}_{\langle ij\rangle_a}$, as shown in Fig.~\ref{FigLattice}. Kitaev Hamiltonian (Eq.~\ref{KitaevHamiltonian}) has a local $Z_2$ gauge invariance in the extended Hilbert space. For practical purposes, the local $Z_2$ gauge transformation amounts to $ u_{\langle ij \rangle a} \rightarrow
\tau_i u_{\langle ij \rangle a} \tau_j$, with $\tau_i =\pm 1$. Eq.~\ref{cons1} is the Gauss law and the physical subspace is the gauge invariant sector.

Eq.~\ref{KitaevHamiltonian}, with conserved ${\hat u}_{\langle ij\rangle_a}$ is the Hamiltonian of free Majorana fermions in the background of frozen $Z_2$ vortices or $\pi$-fluxes. 
Since $Z_2$ gauge fields have no dynamics, all eigenstates can be written as products of a state in the $2^{\frac{1}{2}(4N)}$ dimensional Fock space of  the $c_i$ Majorana
fermions and the $(2)^{\frac{3}{2}(4N)}$ dimensional space of $Z_2$ link variables. We will refer to the former as {\em matter sector} and the 
latter as {\em gauge field sector}. Gauge copies (eigen-states with same energy 
eigen-values) spanning corresponding extended Hilbert space are obtained  by local gauge transformation on  $u_{\langle ij \rangle}$. 

In the gauge field sector we have gauge invariant $Z_2$ vortex charges $\pm 1$ (0 and $\pi$-fluxes), defined as product of $u_{\langle ij \rangle a}$ around each elementary square and octagonal plaquette. These Z$ _{2} $ fluxes (0, $ \pi $) correspond to eigen values (-1, +1) of the plaquette operators S$ _{p} $ and O$ _{p} $.

For a given gauge flux sector, the dimension of the Fock space of the matter sector is 2$ ^{2N} $ rather than 2$ ^{4N} $. This is because we have a hopping Hamiltonian of Majorana fermions rather than complex fermions which has  $2N$ complex fermion oscillators with energy eigen values $ \geq $ 0 within the square Brillouin zone. The size of the BZ is $ \frac{2\pi}{a} $, where $a$ is the periodicity of the lattice.

We have analysed the spectrum of this model in different flux sectors looking for pseudo Fermi surface. We succeeded in finding a Fermi surface in the sector, where each S$ _{p} $ and O$ _{p} $ has eigen value $-1$. This corresponds to Z$ _{2} $ charge of $+1$ (zero flux) in all elementary plaquettes in the gauge field sector. By use of Lieb's theorem we find that the zero flux sector
(or equivalently all S$_{p}$ and O$_{p}=-1$) is not the minimum energy sector. However, we can make it a minimum energy sector by an addition of a chemical potential term $ \lambda \sum_p (S_{p} + O_{p})$, with $\frac{\lambda}{J} >> 1$, which commutes with the Hamiltonian.

\begin{figure}[t]
\begin{center}
\includegraphics[width=3in]{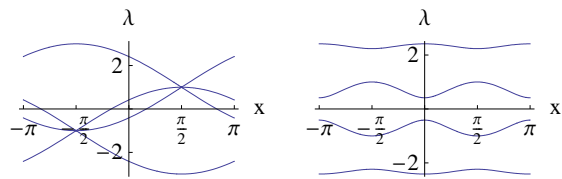}
\caption{Spectrum of $iA(q)$ for$J_x=J_y=J_z=1$ along $(q_1,q_2)=(x,-x)$ (left) and along $(q_1,q_2)=(x,x)$ .}
\label{spectrumAlongALine}	
\end{center}
\end{figure}

\begin{figure}[t]
\begin{center}
\includegraphics[width=3in]{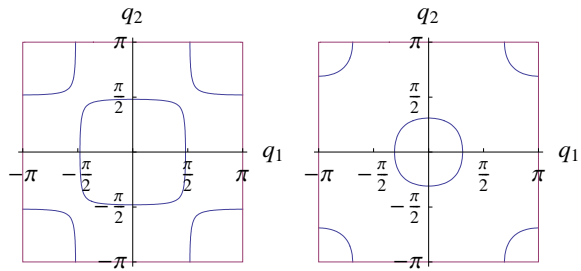}
\caption{dependence of the size and shape of the Fermi surface (in a suitable gauge) on the coupling strengths $J_a$ when all plaquettes have  zero flux. We fix $J_x=J_y=1$ and $J_z$ takes values $0.5$ and $1.9$ respectively for the left and right figures. }
\label{FigJVariation}	
\end{center}
\end{figure}

The gauge is chosen such that $u_{\langle ij \rangle_a}=+1$ if, in Fig.~\ref{FigLattice}(a), the arrow in the bond $\langle ij \rangle_a$ point from $i$ to $j$. The resulting lattice is periodic with each diamond forming the unit cell which is repeated along primitive vectors $\mathbf{n}_1=(1,0)$ and $\mathbf{n}_2=(0,1)$. Using a new notation $c_{i \lambda}$ for the Majorana fermions, where $i$ correspond to the unit cell and $\lambda$ the position within it, the Fourier transformation $a_{\lambda}(q)=\sum_i e^{-i \mathbf{q} \cdot \mathbf{r}_i} c_{i \lambda}/\sqrt{2N}$ transforms the Hamiltonian to
\bearr
H=\frac{i}{2}  \sum_q a^{\dagger}_{\lambda}(q) A_{\lambda,\mu}(q) a_{\mu}(q),
\eearr
with $\{ a_{\lambda}(q),a^{\dagger}_{\mu}(k)\}=\delta_{q,k} \delta_{\lambda,\mu}$,~ $a^{\dagger}_{\lambda}(q)=a_{\lambda}(-q)$ and 
\bearr 
\label{eqA}
iA(q)=2 \left [
\begin{array}{cc}
J_x \sigma^y & -iJ_y \sigma^x+i J_z \alpha\\
iJ_y \sigma^x-iJ_z \alpha^{\dagger} & -J_x \sigma^y
\end{array} \right ],
\eearr
where $\alpha$=diag$(e^{i q_2},-e^{-i q_1})$ and $q_i=\mathbf{q}\cdot \mathbf{n}_i$.  The skew-Hermitian matrix $A(q)$ of size $2n \times 2n$ with $n=2$ has the property $A^T(q)=-A(-q)$. If $\mathbf{v}_i(q)$ is an eigenvector of $iA(q)$ with positive eigenvalues $\epsilon_i(q)$, then $\mathbf{v}^{*}_i(-q)$ is also an eigenvector of $iA(q)$ with eigenvalue $-\epsilon_i(-q)$. Therefore, we could seperate the spectrum of $iA(q)$ into $m(q)$ positive eigenvalues and $2n-m(q)$ negative eigenvalues with $m(-q)=2n-m(q)$. Then, the transformation, $d_i(q)=v^{*}_{ij}(q) a_j(q), ~ d'_{k}(q)=v_{kj}(-q) a_j(q), ~ i \in \{1,..,m(q)\}, ~ k \in \{1,..,2n-m(q)\}$ diagonalises $H$, and the new operators are related by $d^{\dagger}_{i}(q)=d'_{i}(-q)$. The Hamiltonian becomes
\bearr
H&=&\frac{1}{2} \sum_q \sum_{i=1}^{m(q)}  \epsilon_i(q)  \left[ 
d^{\dagger}_{i}(q) d_{i}(q)-d'^{\dagger}_i(-q) d'_i(-q)
\right ] \\
&=&\sum_q \sum_{i=1}^{m(q)}  \epsilon_i(q)  \left[ 
d^{\dagger}_{i}(q) d_{i}(q)-1/2 \right ],
\eearr
with the new operators obeying $\{d_i(q),d^{\dagger}_j(q')\}=\delta_{ij}\delta_{q q'}$. The ground state energy is given by $-\frac{1}{2}\sum_q \sum_{i=1}^{m(q)} \epsilon_i(q)$. For gapless excitations, at least one member of the spectrum should be zero, the condition for which is given as $Det(iA)=0$. It is readily seen that this is possible if the condition $(J_x-J_z)^2 \leq J_y^2 \leq (J_x+J_z)^2$, which becomes $J_x \leq J_y+J_z,~J_y\leq J_x+J_z,~J_z\leq J_x+J_y$  when all $J_a$ are positive, is satisfied.   Fig.~\ref{spectrumAlongALine} shows the eigenvalue spectrum $\epsilon_i(q)$ of $iA(q)$ in Eq.~\ref{eqA} along lines in the $(q_x,q_y)=(x,-x)$ and $(q_x,q_y)=(x,x)$. The points $\epsilon_i(q)=0$ belong to the Fermi surface. One could see that these points are not symmetric about $(0,0)$ in this gauge. To exhibit the variation of size and shape of the Fermi surface with the parameters $J_a$, we use a gauge which has the symmetry of the original lattice and the result is reported in Fig.~\ref{FigJVariation}.  The volume of the Fermi surface becomes zero as we move out of the triangle defining the above mentioned inequality.

To study topological properties of the gapless phase, characterised by Chern numbers, we added 3-spin interaction terms (which is equivalent to adding an external magnetic field). Unlike the Kitaev model on a honeycomb lattice, this does not open a gap. Any non-trivial topological property of the pseudo Fermi sea remains hidden at the moment. The 3-spin interaction,  for the  spins in the central square plaquette given in Fig.~\ref{FigLattice}(a) can be written as 
$K \left( \sigma_1^x \sigma_2^z \sigma_3^y + \sigma_2^y \sigma_3^z \sigma_4^x+\sigma_3^x \sigma_4^z \sigma_1^y+\sigma_4^y \sigma_1^z \sigma_2^x\right)$. 
In terms of the Majorana fermions, the additional term in the Hamiltonian becomes
\bearr
H^{(3)}=\frac{iK}{2}\sum_{jk} Z_{jk} c_j c_k,
\eearr
where the sum is over second-nearest-neighbors and $Z_{jk}=-Z_{kj}$ depends on the chosen guage $u_{ij}$. In fig.~\ref{FigLattice}(b) these interactions are shown as dashed lines and $Z_{ij}=+1$ on the link $\left\langle ij\right\rangle$ if the arrow points from $i$ to $j$. For the second-nearest-neighbors in the square plaquette there are two different 3-spin interactions that involves the same end spins. It turns out that they cancel each other in the gauge we have chosen. For example, in Fig.~\ref{FigLattice}(a), spins $1$ and $3$ interact through $\sigma_1^x \sigma_2^z \sigma_3^y$ and $\sigma_3^x \sigma_4^z \sigma_1^y$ which in the chosen gauge adds upto zero when represented in terms of Majorana fermions. The effect of the 3-spin interactions is to modify $A(q)$ by adding to it the matrix
\bearr
A^{(3)}(q)=2K
\left [
\begin{array}{cccc}
 0 &  \lambda_1(q) & 0 &  \lambda_2(q)\\
-\lambda^*_1(q) & 0 & \lambda_2(q)& 0\\
0 & -\lambda_2^*(q) & 0 & \lambda^*_1(q)\\
-\lambda^*_2(q) & 0 & -\lambda_1(q) & 0
\end{array}
\right ],
\eearr
where $\lambda_1(q)=e^{iq_1}+e^{iq_2}$ and $\lambda_2(q)=e^{-iq_1}+e^{iq_2}$. Fermi surface volume decreases with increasing $K$ and vanishes asymptotically.

Thus we have a Fermi surface as an exact solution of a non-trivial spin-\half model in 2 dimensions. An important issue is the meaning of the area enclosed by the Fermi \textit{surface}. Is there a Luttinger theorem that controls the area enclosed by the Fermi surface. Particularly surprising is the fact that the volume of the FS changes with change in ratio of the coupling constants (Fig.~\ref{FigJVariation}). 

We rationaise our result by comparing the band structure of our problem with the problem of a complex fermion hopping on the same lattice in the same flux sector, the hopping amplitude between nearest neighbors being $J_a$, namely we consider the Hamiltonian 
$
 H=-\sum_{a}J_a\sum_{\left\langle ij \right \rangle_a} c^{\dagger}_i c_j +h.c.
$
 In this case we have four single particle bands, because there are four sites per unit cell. Dimension of the Fock space is 2$ ^{4N} $. The ground state corresponds to filling the bands upto zero energy, because of particle-hole symmetry. We have two hole like and two electron like bands. Two of them cross zero energy giving us two Fermi surfaces, one hole like and other electron like. Since we have two electrons (or two holes) per site in the  state, Luttinger theorem in this case efffectively demands only equality of the fermi area of the hole and electron pockets. Further, for the characteristic polynomials $P_{CF}(q_1,q_2,\lambda)$ of the complex fermion hopping (Hermitian matrix) problem and $P_{MF}(q_1,q_2,\lambda)$ of Majorana fermion hopping (skew-Hermitian matrix) problem, in the gauge we have chosen earlier, are connected by the following relation
$
 P_{CF}\left(q_1,q_2,\lambda\right)=P_{MF}\left(q_1+\frac{\pi}{2},q_2+\frac{3\pi}{2},\lambda\right)
$.
There is also a \textit{two to one} correspondence between the positive energy eigen values of \textit{particle and hole} excitation branches of a regular fermi sea and the positive energy complex fermion excitation branch of the \textit{Majorana fermi sea}. That is, in the complex fermion problem we have degenerate positive energy excitations which are hole and electron branches. In the case of Majorana fermion they get identified, meaning that a complex fermion excitation is its own antiparticle.

The present problem can also be solved using Jordan Wigner transformation, without enlarging the Hilbert space~\cite{FengZhangXiang}. In this case the Jordan Wigner fermions have p-wave nearest neighbor pairing term with certain fixed pairing amplitude. What is interesting is that this pairing does not manage to open even partial gap in the Fermi surface. This is reminescent of the RVB mean field solution reviewed earlier, where an extended-S pairing leads to a Fermi surface. Further, nesting instability is imminent, as we have two Fermi surfaces with identical shapes. This is also very similar to RVB mean field result for square lattice Heisenberg antiferromagnet. Additional perturbation will lead to nesting instabilities. It will be interesting to study this along the lines of work by Mandal, Sengupta and Shankar\cite{MandalSenguptaShankar}. 
Equal time spin spin conrrelation function in this model also continues to vanish beyond nearest neighbors, as expected in all Kitaev models\cite{MandalShankarBaskaran}. So what we have is a non trivial spin liquid with ultra short range spin-spin correlations that supports gapless Fermi surface excitations.

\end{document}